\documentclass{ieeeconf}

\usepackage{lipsum}
\usepackage{graphicx}
\usepackage{textcomp}
\usepackage{stfloats}
\usepackage{url}
\usepackage{verbatim}
\usepackage{graphicx}
\usepackage{fancyvrb}
\usepackage{fvextra}

\usepackage{booktabs}
\usepackage{xcolor}
\usepackage{comment}
\usepackage{lipsum}
\usepackage{balance}
\usepackage{color,listings}
\definecolor{lightgray}{rgb}{.7,.7,.7}
\newsavebox\lstbox

\begin{document}
%
\title{Semantic API Alignment: Linking High-level User Goals to APIs}
%
%
%


\author{Robert Feldt$^{1}$\thanks{$^{1}$Robert Feldt is with the Software Engineering division of Chalmers Institute of Technology, Göteborg, Sweden. Contact: robert.feldt@chalmers.se}, Riccardo Coppola$^{2}$\thanks{$^{2}$Riccardo Coppola is with the Department of Computer and Control Engineering of Politecnico di Torino, Turin, Italy. Contact: riccardo.coppola@polito.it }}

\markboth{11th International Workshop on Artificial Intelligence and Requirements Engineering - AIRE, 2024}%
{Shell \MakeLowercase{\textit{et al.}}: Bare Demo of IEEEtran.cls for IEEE Journals}

\maketitle

\begin{abstract}


Large Language Models (LLMs) are becoming key in automating and assisting various software development tasks, including text-based tasks in requirements engineering but also in coding. Typically, these models are used to automate small portions of existing tasks, but we present a broader vision to span multiple steps from requirements engineering to implementation using existing libraries. This approach, which we call Semantic API Alignment (SEAL), aims to bridge the gap between a user's high-level goals and the specific functions of one or more APIs.

In this position paper, we propose a system architecture where a set of LLM-powered ``agents'' match such high-level objectives with appropriate API calls. This system could facilitate automated programming by finding matching links or, alternatively, explaining mismatches to guide manual intervention or further development.

As an initial pilot, our paper demonstrates this concept by applying LLMs to Goal-Oriented Requirements Engineering (GORE), via sub-goal analysis, for aligning with REST API specifications, specifically through a case study involving a GitHub statistics API. We discuss the potential of our approach to enhance complex tasks in software development and requirements engineering and outline future directions for research.

\end{abstract}


%
\IEEEpeerreviewmaketitle

\section{Introduction}

There has been much recent interest in using generative AI models that can perform high-quality text analysis and generation, e.g. so-called Large Language Models (LLM). LLMs belong to the broader category of Deep Learning models, and address the area of natural language processing, by interpreting and generating human-like text. The nature of many software engineering artefacts, which are in many cases repetitive in their structure, suggests that LLM-based agents could help support both analysis and generation. Some of the currently most popular LLMs, in OpenAI's GPT series of models and interfaced through the ChatGPT application, have been widely used in software engineering research and practice in recent times, e.g. for production or test code and in supporting debugging. 

While most current applications have focused on a single software development task, e.g. debugging, and\slash or used a single or a few interactions with a single LLM instance, there are also studies that proposed the use of multiple LLMs to play diverse roles and interact in a more autonomous fashion to, collectively, support more complex tasks~\cite{feldt2023autonomous,yoon2023autonomous}.

In the field of Requirements Engineering (RE), very recent studies have explored the possibility of extracting goal models from natural language requirement specifications \cite{das2024extracting}, how to extract domain models from textual requirements \cite{arulmohan2023extracting}, or how to increase the quality of use cases with the aid of LLM-based agents \cite{de2023echo}. Again, we see that the focus is on a single task in the context of existing development processes.

In this research preview paper, we conceptualize the application of Large Language Models to the extraction and semantic linking of actor (end user) \emph{goals} in the process of requirements engineering, starting from a high-level description or documentation of a software of interest. 

The benefits of this alignment can be two-fold: (i) the API calls extracted from requirements can be mapped to already implemented endpoints, and (ii) the abstracted API calls can indicate features that are not yet available but maybe should be, i.e. to support design and implementation of the API. In fact, since this semantic API alignment, can be interpreted both as a top-down (from user goals) and a bottom-up activity (from existing APIs) we call it \textit{semantic API alignment} (SEAL).



\section{Background and Related Work}


As our reference framework for the definition of goals, we refer to Goal-Oriented Requirements Engineering (GORE) as introduced by A. van Lamsweerde in 2001 \cite{van2001goal}. The final objective of GORE is the identification of all \emph{goals} of the system, defined as \emph{Objectives that the system under consideration should achieve}. Goals can be formulated at different levels of abstraction, ranging from high-level strategic concerns to low-level technical ones. The main pillars of the GORE technique are the following:

\begin{itemize}
    \item \emph{Goal Modeling}: modeling goals according to intrinsic features (e.g., goal type and goal attributes) and links to other goals or other elements of a requirements model (e.g., actors of the system). 
    \item \emph{Goal Specification}: precise specification of goals to support requirements elaboration. 
    \item \emph{Goal Reasoning}: elaboration of the goal by verifying that they correspond with the requirements of the system; validating the goals by identifying scenarios covered by them, and operationalizes the goals. 
\end{itemize}

Regarding the use of LLM agents for Requirements Engineering, a synthesis of the possible applications of Large Language Models to Software Engineering have been conceptualized in a survey by Fan et al. \cite{fan2023large}. In the paper, it is concluded that LLM has found application in most sub-disciplines of Software Engineering, including coding, design, requirements, repair, refactoring, documentation and analytics. It is however highlighted how many challenges remain, especially related to the reliability of the proposed solutions and the need for hybrid approaches involving human supervision.


Several applications of NLP and LLM have been conceptualized for API testing. Kim et al. have described NLPtoREST \cite{kim2023enhancing}, an approach for the extraction of OpenAPI rules for the generation of REST API scripts. The architecture they propose separates the machine-readable part of the OpenAPI specification from the human-readable summary part, which is fed to an NLP-based rule extraction engine. The same authors also presented RESTGPT \cite{kim2023leveraging}, an approach using GPT-3.5 to extract parameter types, formats, examples and operation types from the human-readable part of the description of OpenAPI specifications.

\section{Vision - SEAL}

\begin{figure}
\centering
    \includegraphics[width=\columnwidth]{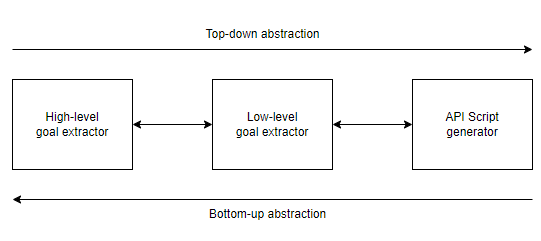}
    \caption{Main components of the conceptual framework. 
    }
    \label{fig:archi}
\end{figure}

\begin{figure}
\centering
    \includegraphics[width=\columnwidth]{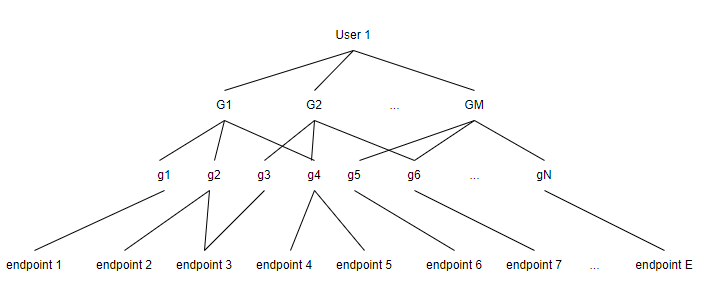}
    \caption{The conceptualized alignment of goals into endpoint calls}
    \label{fig:goals}
\end{figure}

Our vision is to (partly) automate the intermediate steps of what we call \emph{Semantic API Alignment}, i.e., the creation of a model that links high-level user goals or needs, through intermediate sub-goals (possibly organized in hierarchical levels) to the low-level actions that can be implemented through one or more existing software APIs.

This vision is based on existing research that use multiple LLMs with individual tasks and information to collectively accomplish a more complex task~\cite{yoon2023autonomous}. Here, we conceptualize an architecture of LLM-based agents that can operate autonomously to abstract a software description, or of its existing documentation, and devise API calls to obtain the goals related to the functional requirements of the software.

We define a sequence of steps for our conceptual solution that is schematized in Fig. \ref{fig:archi}. The first step is \emph{high-level goal extraction}: the analysis of existing documentation of or relevant for the software, either informally written (e.g., readme documents of software, description of GitHub projects, early design documents or ideas) or in the form of a, more formal but typically still in natural language, requirements document. This step is in charge of performing the \emph{Goal Modeling} and the \emph{Goal Specification} phases of the GORE technique for requirements engineering. Specifically, once an actor (or a set of actors) is identified for the software, the objective is to identify the main goals that they can have when utilizing the software. 

The \emph{low-level goal extraction} step roughly corresponds to the \emph{Goal Reasoning} phase of the GORE technique for requirements engineering. Specifically, in the presence of complex high-level user goals, this step decomposes them into lower-level user goals, possibly in a hierarchy (Fig.~\ref{fig:goals}).

The \emph{API Script generation} step finally performs a mapping of the goals to the existing endpoints of the application. It receives as inputs the set of low-level goals for the software, and the documentation of the APIs of the software. The API documentation can be in the form of a Swagger file or other equivalent documentation. The output of this step of the solution is a set of calls to API endpoints to enforce the low-level goals that can be executed with the set of available endpoints. Of course, if the current API doesn't support some of the goals the system should indicate this and thus allow the developers to either change the goal model or extend\slash re-design the API.

To build the full framework, we envision the construction of an LLM-based multi-agent cognitive architecture already adapted in the field of mobile testing~\cite{feldt2023autonomous,yoon2023autonomous}. Like these existing works, our architecture will need to involve several LLM-based modules to perform also general tasks and support (semi-)autonomous operation, by looping between the agents to iteratively refine the solution, of the system: 

\begin{itemize}
\item \textit{Planner}: creates a step-wise plan of tasks to be achieved when performing the semantic alignment of goals to API calls and vice-versa, and re-plans\slash adapts based on the success\slash failure of recent tasks.
\item \textit{Actor}: chooses the appropriate next action for the next task of the plan, to be achieved, potentially generating and running API calls. 
\item \textit{Observer}: observes the outcomes of the latest Actor actions to judge task success or provide feedback for error correction, in an inner loop with the Actor.
\item \textit{Reflector}: is executed at the end of one whole round of execution of the inner loop, to summarize the overall progress in achieving the high-level goals with the given API. Based on this summary the Planner can decide on further tasks and run the inner, action loop again to further refine results.
\end{itemize}


While the current architecture is based on top-down goal refinement this can also be complemented with a set of agents doing bottom-up analysis and API execution, much like how the mobile testing framework executes actions on a running app~\cite{yoon2023autonomous}.
In the following, we provide a running inspirational example, through interaction with the chatGPT agent, of the \emph{top-down} approach.

\section{Inspirational Example}

To verify the feasibility of the approach, we performed a set of conversational interactions with the GPT-4 engine, by using the online ChatGPT interface. The full prompts that were used are reported in a digital appendix \footnote{doi.org/10.6084/m9.figshare.25201019}. We manually fed the prompt to the GPT-4 engine to mimic the autonomous actions that will be performed by each individual agent.

As a case study, we selected one of the enterprise applications in the EMB database\footnote{https://github.com/EMResearch/EMB}, already used for RESTful API automated test case generation. We selected an application that had a general high-level description to be fed as input to the high-level goal abstractor, and a Swagger API description with textual documentation. After performing that filtering, we selected CatWatch as it has an average number of endpoints in the set (24 REST endpoints)\footnote{https://github.com/zalando-incubator/catwatch}.

Prompt P1 asks the conversational agent to provide high-level goals for the application from the point of view of an individual actor. Since CatWatch is an application to monitor GitHub repositories, we considered as an actor the owner of a GitHub account. \\

{
\tiny

\begin{Verbatim}[frame=single, breaklines=true]
**Prompt: "Elicit High-Level Goals for a Specific Stakeholder in a New Software Project"**

*Context:* You are assisting in the goal elicitation process for: [CatWatch is a web application that fetches GitHub statistics for your GitHub accounts, processes and saves your GitHub data in a database, then makes the data available via a REST API. The data reveals the popularity of your open source projects, most active contributors, and other interesting points. As an example, you can see the data at work behind the Zalando Open Source page. To compare it to CoderStats: CatWatch aggregates your statistics over a list of GitHub accounts.].

*Stakeholder Description:* [Owner of a GitHub account]

*Task:* Based on your understanding of the typical needs and interests of this specific stakeholder in such projects, help generate a list of high-level goals.
\end{Verbatim}

}

The second prompt Prompt P2 asks the conversational agent to decompose the high-level goals provided as the output of the previous prompt into a set of low-level goals. The prompt has to be repeated multiple times (for each of the high-level goals elicited).\\

{
\tiny

\begin{Verbatim}[frame=single, breaklines=true]
**Prompt: "Elicit Low-Level Goals for a Specific Stakeholder in a New Software Project"**

*Context:* You are assisting in the goal refinement process for a software. The high-level goals of the software are the following:

[RESULT OF PROMPT 1]

*Task:* Based on your understanding of the typical tasks that compose the sequence of high-level goal, provide if possible a decomposition of goals into sub-goals. Each low-level goal should theoretically correspond to a single action of the actor with the software.
\end{Verbatim}
}

Prompt P3 reads the JSON representation of the API and providing a list of endpoints along with the necessary details to call them. 
The output of the used prompt is, for each endpoint: the name of the endpoint; a textual description of the endpoint; the available verbs on the endpoint; the parameters for the endpoint (type and name); the output type of the endpoint and the provided result. \\

{
\tiny

\begin{Verbatim}[frame=single,breaklines=true]
**Prompt: “Extract API documentation from a Swagger FILE”**

*Context:* You are assisting in the API abstraction of a given software. The swagger file documenting the APIs of the software is the following:

[SWAGGER FILE]

*Task*: please list all the endpoints from this swagger file excerpt and list for each of them:
-	the path of the endpoint
-	the verb used for the endpoint
-	the tag of the endpoint
-	the summary of the endpoint
-	the description of the endpoint
-	the operationId of the endpoint
-	the consumed and produced type of the endpoint
-	the parameters to call the endpoint.
\end{Verbatim}
}

Prompt P4 receives as input the list of low-level goals (output of P2) and the list of REST endpoints (output of P3). The result of the prompt is a 1-to-many mapping between the low-level goals and API calls.\\ 

{
\tiny

\begin{Verbatim}[frame=single,breaklines=true]
**PROMPT: “Mapping between API endpoints and low-level goals” **

*Context:* You are assisting in the mapping of goals of a software to its API endpoints.
The low-level goals of the software are the following: 

[RESULTS OF PROMPT 2]

The endpoints of the software are the following:

[RESULTS OF PROMPT 3]

*Task*: please map each goal to a sequence of calls to the given endpoints. If a goal is not mappable to an endpoint, please write that the goal cannot be enforced by using the current set of endpoints.
\end{Verbatim}
}

\begin{table}
\tiny
\centering
\caption{High-level goals (hlGs) for CatWatch}
    \begin{tabular}{lp{2cm}p{5cm}}
    \toprule
    hlG n. & hlG name & hlG description\\
    \midrule
    1	& Monitor Open Source Project Popularity	& The stakeholder aims to effortlessly monitor the popularity metrics of their open source projects across various GitHub accounts using CatWatch.\\
2	& Identify Key Contributors and Collaborators
& The stakeholder seeks a feature that highlights the most active contributors and collaborators in their GitHub repositories through CatWatch.\\
3	&Receive Timely Notifications on Project Activity	&The stakeholder desires a notification system within CatWatch that alerts them promptly about significant activities, such as new contributions or rising project trends.\\
4	&Ensure Data Security and Privacy Compliance &	The stakeholder insists on CatWatch implementing robust data security measures and compliance with privacy standards to safeguard their GitHub account information.\\
5	&Seamless Integration with Existing Workflow Tools
&The stakeholder requires CatWatch to seamlessly integrate with their existing workflow tools and development environments, enhancing productivity and user experience.\\
6	&Access Comprehensive Analytics and Reports &	The stakeholder aims to access detailed analytics and reports generated by CatWatch, offering insights into project performance, community engagement, and other relevant metrics.\\

    \bottomrule
    \end{tabular}
    \label{tab:resprompt1}
\end{table}

\begin{table*}
\tiny
\caption{Low-level goals (llGs) for CatWatch }
        \begin{tabular}{lp{3.5cm}p{9cm}p{3cm}}
    \toprule
    hl.ll G & llG name & llG description & Mapping to endpoints\\
    \midrule
1.1	& Check Popularity Now & Quickly view how popular my open source projects are right now using CatWatch.
 & /projects, /statistics/projects \\
1.2	&Explore Project Popularity Over Time & 	Dive into historical trends and see how popular my open source projects have been using CatWatch. & /projects, /statistics/projects\\
2.1	&Identify and Connect with Top Contributors	&Find and connect with the most active contributors to my GitHub repositories through CatWatch. &  /contributors \\
2.2&	Acknowledge Collaborators' Contributions	&Use CatWatch to highlight the impact and significance of collaborators' contributions, fostering a sense of community. & /contributors \\
3.1 & Customize My Notification Preferences & Personalize my notification settings in CatWatch to receive alerts tailored to specific project activities. (Not applicable with current set of APIs) & (Not applicable with current set of APIs)\\
3.2 & Get Instant Notifications on Key Events & Receive real-time notifications in CatWatch about new contributions or significant trends in my GitHub repositories. (Not applicable with current set of APIs) & /config/scoring.project\\
4.1 & Ensure My Data is Encrypted & Trust CatWatch to implement robust data encryption methods, securing my GitHub account information from unauthorized access. & (Not applicable with current set of APIs)\\
4.2 & Control Who Accesses My Data & Use access controls in CatWatch to manage and restrict access to my GitHub account information, ensuring only authorized personnel can interact with it. & (Not applicable with current set of APIs)\\
5.1 & Easily Integrate CatWatch with My Workflow & Benefit from CatWatch's robust integration framework to seamlessly connect the application with my existing workflow tools and development environments. & (Not applicable with current set of APIs)\\
5.2 & Follow Integration Documentation for a Smooth Process & Refer to comprehensive integration documentation within CatWatch to guide me on effectively integrating the application with my established workflow. & (Not applicable with current set of APIs)\\
6.1 & Explore Detailed Analytics Reports & Utilize CatWatch to generate and explore detailed analytics reports, gaining insights into project performance, community engagement, and relevant metrics. & /statistics/projects\\
6.2 & Customize Report Parameters & Tailor analytics reports in CatWatch by customizing parameters, allowing me to focus on specific project performance and engagement metrics of interest. & /statistics/projects\\
\bottomrule
\end{tabular}
\label{tab:resprompt2}
\end{table*}

The results of prompt 1 are provided in table \ref{tab:resprompt1}. For a GitHub project curator, the LLM agent identifies six main high-level goals for the CatWatch project.

From the results, we notice that some of the high-level goals are indeed non-functional (e.g., hlG 3 and hlG 4). These goals should be verified by an additional self-critique component of the architecture to be cut out from the results instead of being considered for the alignment of API calls. 

The low-level goal names genereated by P2, along with a one-line description, are reported in the first column of table \ref{tab:resprompt2}. The generation of low-level goals required multiple interactions with the OpenAI chat engine to tailor the responses provided by the editor. Most of the low-level goals provided had initially to be discarded since they were given from a development perspective and not from the perspective of the final user of the application. 

In the last column of table \ref{tab:resprompt2} we report the results of Prompt 4, which performed the mapping from low-level goals to (sequences of) endpoint calls. After several calibration interventions performed by interactions with the chat interface, the agent was able to map 7 out of 12 low-level goals to API calls. The non-functional goals were correctly identified as not mappable to existing endpoints. It is worth mentioning that, even without providing details about the exact input and output to be provided and to be expected, the model identified sequences of API calls for some of the low-level stakeholder's goals.

\section{Progress towards vision}

The preliminary interactions performed with the LLM agents provided a set of high-level goals, then refined into lower-level goals and eventually mapped into sequences of API calls described in the swagger file of a simple web application. The repeated interactions that had to be performed with the agents made it clear that such alignment of high-level documentation and software description cannot easily be performed. A general task setting architecture with self-critique and both an inner and an outer loop, with agents dedicated to the creation of high- and low-level goals and others dedicated to the analysis and refinement of their output would be needed. 

The results provided by the OpenAI GPT-based agents also suffer from non-homogeneous size and precision with the provided input: in many interactions, some of the provided high- or low-level goals were discarded from the output with no reasoning provided in the output. This additional limitation is another incentive for the inclusion of dedicated agents in charge of performing simple controls in the output of other elements of the architecture. 

The final output of our inspirational example was the identification of one or more endpoints to call to pursue a given low-level goal. It is, of course, necessary to infer additional details to understand the type of data to provide as input to the given endpoint, and the way the output of the endpoints has to be consumed and fed to the next ones. 

It is however worth underlining that generating a plausible sequence of API calls might remain an unfeasible task when the endpoints to enforce are strictly tied to business processes or protocols that are very specific to a company or domain, and are not easily retrievable from other existing documented applications or the knowledge of the GPT agents. 

\section{Conclusion}

In this work, we proposed -- through a conceptual architecture and an inspirational running example -- the application of a system of multiple LLM agents for the semantic alignment of API calls: bridging from high-level user goals to concrete API endpoints. We argue that this type of application of LLMs in requirements and, generally, in software engineering is likely to be long-term useful, i.e. spanning multiple steps of current, manual development processes. Based on our manual use of a LLM to perform several of the steps involved we saw that a more complex architecture that can partly divide up its work and act on it semi-autonomous is likely needed. To enable human guidance in this process will likely also be fruitful. Thus, our proposal is but one instance of a more general future for requirements (and software) engineering, that we envisage, where systems of multiple LLM agents interactively can augment the work of software engineers. Existing requirements and software engineering methods, such as goal-oriented requirements engineering, can act as concrete implementation guidance and the flexibility of LLMs actually be used to ``implement'' and ``execute'' them.

As our immediate future work, we plan the implementation and empirical evaluation of the architecture with state-of-the-art conversational agents.

\bibliographystyle{IEEEtran}
\bibliography{bib}

\end{document}